\documentclass[11pt,a4paper,oneside]{article}
\usepackage[T1]{fontenc}
\usepackage[utf8]{inputenc}
\usepackage{amsmath,amsthm,amsfonts,amssymb,makeidx,mathtools}
\usepackage[english]{babel}
\usepackage{amsxtra,amscd}
\usepackage{cancel,nicefrac}
\usepackage{bigints}
\usepackage{appendix}
\usepackage{enumitem}
\usepackage{xcolor}

\newtheorem{Proposition}{Proposition}
\newtheorem{Axiom}{Axiom}

\def\@{\hskip.8pt}
\def\?{\hskip.3pt}

\def\V#1{{\mathcal V}_{#1}}
\def\a{\alpha}
\def\b{\beta}
\def\g{\gamma}

\def\p{\mathfrak q}

\def\th{\vartheta}
\def\vphi{\varphi}

\def\Chi{\kappa}

\def\Frac#1#2{{\textstyle\lower.8pt\hbox{\Large$\frac{#1}{#2}$}}}

\def\d#1/d#2{\frac{d#1}{d#2}}                                                   
\def\D#1/d#2{\Frac{d#1}{d#2}}                                                   
\def\dD#1/d#2{{\textstyle{\text{\Large$\frac{D\?#1}{D\?#2}$}}}}                 

\def\de#1/de#2{\frac{\partial#1}{\partial#2}}                                   
\def\De#1/de#2{\Frac{\partial#1}{\partial#2}}                                   

\def\SD#1/de#2/de#3{\ifx#2 \frac{\plus02\partial^{\@\@2}#1}
    {\plus90\partial\@#3^{\@2}} \else\frac{\plus02\partial^{\@\@2}#1}
    {\partial\@#2\@\partial\@#3}\fi}

\def\DD#1{\tilde\partial\?_#1}

\def\Per{\otimes}

\def\var{\operatorname{var.}}

\def\const{\text{const.}}

\def\|{;\@}

\title{A Machian Approach to Relativistic Cosmology}

\author{%
Enrico Massa\thanks{DIME, Sez.~Metodi e Modelli Matematici, Università di Genova, Via All'Opera Pia 15, 16145 Genova, Italy. Email: \texttt{massa@dima.unige.it}}
\and
Davide Astesiano\thanks{Science Institute, University of Iceland, Dunhaga 3, 107 Reykjav\'{\i}k, Iceland. \newline
Mathematics Division, Venturi Space, Route du Pâqui 1, 1720 Corminboeuf, Switzerland. Email: \texttt{davide.astesiano@venturilab.ch}}
}

\date{\today}

\begin{document}
\maketitle

\begin{abstract}
We present a consistent relativistic formulation of Mach’s principle within a geometric theory of gravitation. In this approach, neither inertia nor free fall is 
assumed a priori; instead, the motion of any local system arises from its dynamical interdependence --- direct or indirect --- with the rest of the Universe. This 
viewpoint provides a sharp criterion for what qualifies as a genuinely Machian gravitational theory: a theory is Machian only if the global mass–energy distribution 
underwrites the very existence of the pseudo-Riemannian structure of spacetime, rather than merely determining local features such as curvature on a pre-existing 
background. We analyze the resulting cosmological model and discuss its phenomenological implications. In the appropriate local limit, the theory reduces to General 
Relativity, thereby preserving agreement with all Solar System tests. 
\end{abstract}

\section*{Introduction}\label{S0}
Every theory of space, time, and gravitation rests on a foundational set of primitive concepts --- such as \emph{event, particle, light ray} --- along with a 
corresponding set of axioms specifying how these concepts are represented within an appropriate mathematical framework. 

In what follows, we shall restrict our attention to the class of theories based on the following basic assumption:
\begin{Axiom}[Riemannian hypothesis]\label{Ax1}
The set of events forms a 4-dimensional pseudo-Riemannian manifold of normal hyperbolic type $\V4$, referred to as \emph{space-time\/}. Light rays and freely falling 
test particles are represented, respectively, by null and time-like geodesics in $\V4$. Ordinary particles are described by time-like curves in $\V4$. 
\end{Axiom}

According to Axiom \ref{Ax1}, the metric of $\V4$ is described by a symmetric fundamental form 
\begin{equation*}
\Phi=g_{ij}\,dx^i\otimes dx^j
\end{equation*}
with signature $-+++$\, 
\footnote%
{Latin indices run from 0 to 3. Greek indices run from 1 to 3. Einstein’s summation convention is used throughout. Ordinary derivatives are denoted by a comma, covariant 
derivatives by a semicolon.}.

The aim of a relativistic theory of gravitation is to relate the components $g_{ij}$ to the mass-energy distribution in the physical world by means of suitable field 
equations. 

Upon closer reflection, the problem proves to be tied to a more fundamental issue: the physical characterization of free fall.

It is precisely at this point that Mach’s principle acquires a precise geometrical role. To clarify this, recall that, according to the Machian viewpoint, every local 
system interacts with all other physical systems in the Universe. 

In particular, the total dynamical effect acting on a point-like test particle~$\p$ --- which, in the Machian context, encompasses inertia, gravitation, and all other 
local interactions --- depends explicitly on the attributes of $\p$ (inertial mass, electric charge, etc.), on the state of the ``external universe'' $\mathcal U$ (i.e., 
the totality of sources interacting with $\p$), as well as on the mutual relations between $\p$ and $\mathcal U$. 

The evolution of $\p$ is then determined by the requirement that, at each instant, the dynamical effects acting on it balance one another; that is, by the symbolic 
equation 
\begin{equation}\label{0.1}
\text{total dynamical effect on }\p\@=\@0
\end{equation}
supplemented with suitable initial conditions. 

Within the Machian framework, the goal of particle dynamics is therefore to provide a satisfactory functional expression for the left-hand side of eq.~\eqref{0.1} in any 
physically relevant context. 

In particular, what is commonly referred to as \emph{free fall\/} is nothing but a special case of this general framework, in which all interactions other than inertia 
and gravitation are considered negligible. In this case, by virtue of the equivalence principle, the total dynamical effect on $\p$ is proportional to its inertial mass 
$m_\p$. Accordingly, the left-hand side of eq.~\eqref{0.1} factorizes as the product of $m_\p$ and a field $\psi$, determined by the entire mass-energy distribution in 
the physical world. 

With this in mind, let us now return to Axiom \ref{Ax1}. Here, we are presented with a completely different view of free fall, understood as being entirely determined by 
the pseudo-Riemannian structure of $\V4$ through the geodesic equation 
\footnote%
{This includes light rays, interpreted as the world-lines of freely falling photons.} 
\begin{equation*}
\frac{d^2\@x^i}{d\xi^2}\@+\@\Gamma_{pq}{}^i\,\frac{dx^p}{d\xi}\,\frac{dx^q}{d\xi}\@=\@0
\end{equation*}

Such a formulation can be reconciled with the requirements of Mach’s principle only by assuming that the Riemannian hypothesis tacitly encodes the presence of the 
Universe as a whole --- that is, by envisaging a space-time model in which the \emph{entire Riemannian structure} of $\V4$, rather than merely a local feature such as 
curvature, is determined to the overall mass-energy distribution. 

Any theory consistent with this viewpoint will be classified as a \emph{Machian theory of gravitation}, while theories that are not will be termed \emph{absolute 
space-time theories}. 

According to this classification, General Relativity, in its standard formulation, is an absolute space-time theory, remarkably consistent with experimental data and 
providing a highly accurate, internally coherent description of gravitational phenomena. 

At first glance, this might suggest that the absolute space-time model is closer to physical reality than the Machian one. If this standpoint is accepted (see, among 
others, Ref.~\cite{Synge}), then one is left with no choice but to dismiss Mach’s principle as unphysical and to come to terms with the Newtonian idea that motion is an 
absolute concept. 

At the same time, however, it is difficult to overlook the fact that the Machian view of dynamics --- in which all physical phenomena are ultimately reduced to mutual 
interactions among material bodies --- possesses a greater intuitive appeal than the Newtonian picture. 

The challenge is therefore to determine whether a suitable reinterpretation might allow one to reconcile the use of Einstein’s equations with the requirements of a 
Machian theory of gravitation.

\section{The field equations}\label{S1}

\subsection{Preliminaries}\label{S1_1}
As pointed out in the Introduction, according to the Machian viewpoint, the notion of free fall presupposes the existence of a background relative to which such motion 
is defined. 

General Relativity, formulated as a local theory, naturally fits into this scheme: the background is effectively provided by the ensemble of distant masses in the 
Universe, which play a role analogous to Newton’s ``fixed stars''. 

In this framework, the distant masses are not explicitly referenced; their influence is formally encoded in the pseudo-Riemannian structure assigned to the space-time 
manifold $\V4$, with free fall identified with geodesic motion. Gravitation then manifests itself through geodesic deviation, with local masses acting as sources of 
curvature via Einstein’s equations. 

This model is fully consistent with experimental evidence and remains the most successful theory of gravitation currently available. 

The situation changes radically when the ``system'' under investigation is the Universe itself: in cosmology, there is no external environment left to provide the 
background needed to define inertial motion. 
The origin of the relevant space-time structures must therefore be traced back to the Universe itself, i.e., to the very object under study. 

This motivates the Machian standpoint adopted here: the material content of the Universe is not primarily responsible for \emph{differential} gravitational effects on 
top of a pre-existing arena, but for the \emph{existence and global form} of the pseudo-Riemannian structure of~$\V4$. 

In particular, the explicit form of the metric tensor is taken to depend essentially on the way in which matter is \emph{distributed\/}, with gravitation associated with 
inhomogeneity, and the homogeneous and isotropic limit corresponding to a maximally symmetric (constant-curvature) geometry 
\footnote%
{A more austere approach would associate the absence of gravitation with flatness. The perspective adopted here does not rule out the possibility that, beyond inertia, 
the material background may also give rise to other global effects --- such as those related to the accelerated or decelerated expansion of the Universe ---  commonly 
attributed to dark energy.}. 

This perspective is fully consistent with Einstein’s equations
\begin{equation}\label{1.2}
G_{ij}+\Lambda\@g_{ij}= \Chi\,T_{ij}\,,
\end{equation}
provided that the tensor on the right-hand side is interpreted as encoding the mechanical content of the \emph{inhomogeneities} in the actual material distribution, 
rather than the total background content of the Universe. 

This reinterpretation entails no significant modification to the local applications of General Relativity.

Consider, for example, Schwarzschild's exterior solution. According to the traditional interpretation, it describes the gravitational field surrounding a spherical 
distribution of matter embedded in an otherwise empty Universe. 
In the Machian framework, it describes the field surrounding a spherical \emph{concentration\/} of matter within an otherwise \emph{smoothed out\/} Universe. 

Although conceptually profound, the difference has negligible impact on the solution procedure: in both cases, the metric takes the form
\begin{equation}\label{1.3}
\Phi=-\left(1-\frac{2m}r\right)\@dt\Per dt\@+\@\frac{dr\Per dr}{1-\Frac{2m}r}\@+\@r^2\left(d\th\Per d\th+\sin^2\th\,d\vphi\Per d\vphi\right)\!,
\end{equation}
with $m$ interpreted, in the former case, as the total mass of the spherical distribution, and in the latter as the excess mass due to the concentration. 

In both instances, the essential point is that eq.~\eqref{1.3} provides a highly accurate description of the phenomena occurring, for example, within the Solar System. 
This suggests that, in local problems, the influence of distant celestial bodies --- the so-called ``fixed stars'' --- can, to a good approximation, be ignored. 

In an absolute space-time theory, this is justified by asserting that the contribution of the fixed stars to local phenomena is negligibly small, so that approximating 
the Sun as an isolated mass in an otherwise empty Universe yields a faithful representation of physical reality \cite{Synge}. 

From the Machian standpoint, on the other hand, it suffices to assume that the inhomogeneity of the background Universe is irrelevant to the computation of local 
effects, without excluding the possibility that the very existence of the fixed stars may play a significant role. 

The considerations above are clearly independent of the specific solution under analysis. We may therefore formulate the following:

\begin{Proposition}\label{Pro1_1}
In local applications of General Relativity, both the absolute and Machian approaches lead to the same conclusions. Hence, any choice between them can only be justified 
on cosmological grounds. 
\end{Proposition}

\subsection{The equations}\label{S1_2}      
(i) In the Machian perspective discussed in Subsec.~\ref{S1_1}, the application of Einstein's equations \eqref{1.2} to the smoothed-out Universe yields the trivial 
identity $0=0$. 

This raises important questions concerning such notions as the energy-momentum tensor, the evolution equations, and the very idea of determinism. 

These notions are inherently tailored to a local description of physical reality; in particular, momentum and energy --- along with their associated evolution laws --- 
are meaningful only within a theoretical framework that implicitly presupposes the existence of an external background for reference. 

While this does not rule out the possibility of describing the cosmic matter distribution by means of a tensor $\th_{ij}$, it does preclude interpreting $\th_{ij}$ as an 
energy-momentum tensor in the proper sense. 

What is meaningful, rather, is to assign energy and momentum --- and therefore also an energy-momentum tensor $T_{ij}$ --- to the inhomogeneities present in the actual 
Universe, and to analyse the corresponding evolution laws. 

Needless to say, this is perfectly consistent with the Machian interpretation of the right-hand side of Einstein’s equations \eqref{1.2}. 
These automatically entail the evolution equation 
\begin{equation}\label{1.4}
\left(\Chi\,T^{ij}\right)_{\|j}\@=\@0
\end{equation}

A few comments are in order concerning eq.~\eqref{1.4}:
\begin{itemize}
\item 
In general, the tensor $T_{ij}$ depends on the full set of attributes involved in the description of the inhomogeneities. To establish a well-posed dynamical 
framework, this set must be reduced to four independent variables (typically, the density and four-velocity) via suitable equations of state.

\item
In eqs.~\eqref{1.2} and \eqref{1.4}, the coupling coefficient $\Chi$ is regarded as a scalar field depending on the state of the Universe, rather than as a constant. 
To ensure determinism, eq.~\eqref{1.2} must therefore be supplemented by an additional scalar equation. The motivation for this requirement is cosmological in nature, 
and will be discussed in Sec.~\ref{S1_4}. 

\item
Even for non-constant $\Chi$, a classical argument \cite{Lichnerowicz,Synge} shows that eqs.~\eqref{1.4}, together with Lichnerowicz's matching conditions, imply 
geodesic motion for freely falling point-like inhomogeneities. 
The Machian perspective is thus consistent with the Riemannian hypothesis stated in Axiom~\ref{Ax1}. What breaks down is the conservation law for the proper mass of 
the inhomogeneities. 
This is hardly surprising: in an expanding Universe, the inertial attributes of a given body are naturally expected to depend on the large-scale distribution of all 
other bodies. 

\end{itemize} 

\medskip\noindent
(ii) \,Let us now consider the possibility of establishing a relation between the material content of the actual Universe and the geometry of space-time. 

One possible approach is to include among the attributes of $\V4$ the ``footprints'' left by the reference background underlying the definition of free-fall motion.
To this end, alongside the pseudo-Riemannian structure, we endow $\V4$ with a three-dimensional, completely integrable distribution $\mathcal D$ spanned by space-like 
vectors. 

We interpret the integral surfaces of $\mathcal D$ as representing how the simultaneity of events in the Universe's co-moving frame would appear in the absence of 
inhomogeneities --- that is, as a picture of the one-parameter family of three-spaces associated with the rest frame of the smoothed-out Universe. In this view, the 
future-directed unit vector field $\DD0 = V^i\@\De{}/de{x^i}$\vspace{2pt}, orthogonal to the foliation induced by $\mathcal D$, is naturally identified with the 
four-velocity field of the material substratum. \vspace{2pt} 

The objective is to establish a relation between a left-hand side expressed in terms of the fields~$\Phi$ and~$\DD0$, and a right-hand side encoding the material 
distribution in the actual Universe. 
Such a choice is inherently a creative act, whose validity can be assessed only by examining its implications.  
Our proposal is to adopt a relation of the form 
\begin{equation}\label{1.5}
W_{ij}=\Chi\,\th_{ij}
\end{equation}
where~$\th_{ij}$ represents the overall matter tensor, while $W_{ij}$ is defined as 
\begin{equation}\label{1.6}
W_{ij}=G_{ij}+\Lambda\@g_{ij}-\DD0\big(V^p_{\;\;\|p}\big)\Big[V_i\@V_j+\a\big(V_{i;j}+V_{j;i}\big)\Big]
\end{equation}
with $\a$ denoting a suitable constant.

Let us examine what kind of scenario emerges from equations~\eqref{1.5},~\eqref{1.6}. At first glance, one might expect that, by supplementing the system with an 
appropriate set of constitutive relations, it would be possible to construct a deterministic scheme capable of simultaneously determining the geometry of~$\V4$ and the 
evolution of the cosmos via a well-posed Cauchy problem. 

However, as previously noted, the very notion of evolution becomes meaningful only with reference to some external background. From this perspective, since the Universe 
encompasses the totality of what exists, it can evolve only with respect to itself --- a circumstance that renders the concept inherently ambiguous. 

By contrast, it is perfectly meaningful to investigate the evolution of possible inhomogeneities in the matter distribution. This is achieved by assigning to such 
inhomogeneities an appropriate energy-momentum tensor~$T_{ij}$ --- logically distinct from $\th_{ij}$ --- and by resorting to eqs.~\eqref{1.2} and \eqref{1.4}, possibly 
supplemented by suitable constitutive equations. 

This viewpoint is naturally recovered by recasting Eq.~\eqref{1.5} in the form 
\begin{equation}\label{1.7}
G_{ij}+\Lambda\@g_{ij} = \Chi\biggl\{\th_{ij} +\frac{\DD0\big(V^p_{\;\;\|p}\big)}{\Chi}\Big[V_i\@V_j+\a\big(V_{i;j}+V_{j;i}\big)\Big]\biggr\}\coloneq \Chi\,T_{ij}
\end{equation}
and interpreting the tensor on the right-hand side as describing the deviation of the actual matter distribution from the smoothed-out one --- that is, as the 
energy-momentum tensor associated with inhomogeneities. 

In this approach, the vector field $\DD0$ functions as a \emph{hidden variable\/}. The explicitly accounted features are the inhomogeneities in the matter distribution, 
the coupling coefficient $\Chi$, and the pseudo-Riemannian structure of space-time, construed as the formal imprint of the Universe’s existence.

As already noted, to ensure determinacy, eq.~\eqref{1.7} --- and thus also the original equation \eqref{1.5} --- must be supplemented by a scalar equation 
relating~$\Chi$ to the state of the background Universe.

\subsection{Cosmology}\label{S1_3}
(i) \,Within the framework defined by equations~\eqref{1.5} and~\eqref{1.6}, cosmology assumes a distinctive role, since it concerns a material continuum --- the 
smoothed-out Universe --- whose four-velocity field is regarded as one of the geometric attributes of the manifold~$\V4$. 

As already anticipated, a Machian approach to this setting unfolds along the following lines: 
\begin{itemize}
\item
the Riemann tensor of $\V4$ takes the special form 
\begin{equation}\label{1.8}
R_{ijkl}=\frac\Lambda 3\big(g_{ik}\@g_{jl}-g_{il}\@g_{jk}\big);
\end{equation}

\item
the matter distribution is described by a symmetric, spatially isotropic tensor which, in natural units, reads 
\begin{equation}\label{1.9}
\th_{ij}\@=\@\mu\@V_i\@V_j\@+\@p\left(g_{ij}+V_i\@V_j\right)
\end{equation}
with $V^i\@\De{}/de{x^i}$ identified with the vector field $\DD0\@$;

\item
the field equation \eqref{1.5} must be supplemented by a scalar relation, providing insight into the behaviour of the coupling coefficient~$\Chi\@$.
\end{itemize}

Regarding the last point, we assume the conservation of the material content of the substratum, expressed by the continuity equation
\begin{equation}\label{1.10}
\big(\mu\@V^i\big)_{\|i}\@=\@\DD0(\mu)\@+\mu\@V^i{}_{\|i}\@=\@0
\end{equation}

\medskip\noindent
(ii) \,In view of eqs.~\eqref{1.8}, \eqref{1.9}, eqs.~\eqref{1.5}, \eqref{1.6} yield the relation 
\begin{equation}\label{1.11}
-\@\DD0\big(V^p_{\;\;\|p}\big)\big[\@V_i\@V_j+\a\left(V_{i;j}+V_{j;i}\right)\big]=\Chi\big[\@\mu\@V_i\@V_j\@+\@p\left(g_{ij}+V_i\@V_j\right)\big]
\end{equation}
which is mathematically equivalent to the following three separate equations:
\begin{subequations}\label{1.12}
\begin{align}
& \Chi\@\mu\@=-\@\DD0\big(V^p_{\;\;\|p}\big)\@,                                                         \\[3pt]
& \Chi\@p\@=-\frac23\,\a\@V^p_{\;\;\|p}\,\DD0\big(V^p_{\;\;\|p}\big),                                   \\[3pt]
& V_{i;j}+V_{j;i}\@=\frac23\,V^p_{\;\;\|p}\,\big(\@g_{ij}+V_i\@V_j\big)
\end{align}
\end{subequations}

By virtue of eq.~(\ref{1.12}c), the congruence $\Gamma$ of stream lines of the field $\DD0$ satisfies $V_{i;j}\@V^j=0$, and is therefore both normal (by hypothesis) and 
\emph{geodesic}. As such, it consists of coordinate lines $x^0 = \var$ in a Gaussian coordinate system. 

Recalling that, along each hypersurface $x^0=\const$, the pull-back of the tensor $\Omega_{ij}\coloneq\frac12\big(V_{i;j}+V_{j;i}\big)$ represents the second fundamental 
form, from Gauss' theorem and eqs.~\eqref{1.8}, (\ref{1.12}c) we conclude that the curvature tensor $R^*{}_{\!\a\b\g\delta}$ of each such hypersurface satisfies the 
relation 
\begin{equation*}
R^*{}_{\!\a\b\g\delta}=R_{\a\b\g\delta}-\big(\Omega_{\a\g}\@\Omega_{\b\delta}- \Omega_{\a\delta}\@\Omega_{\b\g}\big)=
\biggl(\frac\Lambda3\@-\@\frac{\big(V^p{}_{\!\|p}\big)^2}9\biggr)\big(g_{\a\g}\@g_{\b\delta}- g_{\a\delta}\@g_{\b\g}\big)
\end{equation*}

In the absence of inhomogeneities, each leaf of the foliation determined by the distribution~$\mathcal D$ therefore has constant curvature.

Summing up and performing standard calculations, we conclude 
\begin{Proposition}\label{Pro1_1}
In the co-moving coordinate system associated with the substratum, the metric takes the Friedmann-Robertson-Walker form 
\begin{equation}\label{1.13}
\Phi=-\@ dt \Per dt+ \@R^2(t)\,\frac{du\Per du+u^2 \left(d\theta \Per d\theta+ \sin^2\theta\, d\varphi \Per d\varphi\right)}{\left(1+\frac k 4\@u^2 \right)^2},
\end{equation}
with $k=1,0,-1$, and with the function $R\/(t)$ obeying the conditions
\begin{subequations}\label{1.14}
\begin{align}
& \ddot R\@=\@ \frac\Lambda 3\,R                                                                         \\[3pt]
& \frac\Lambda 3\,R^2-\dot R^2=R \ddot{R}-\dot{R}^2=\@k,\@,
\end{align}
\end{subequations}
required in order to ensure the validity of eq.~\eqref{1.8}
\end{Proposition}

For any $\g\in\Gamma$, the equation of geodesic deviation, applied to an arbitrary spatial connection vector $X$ along $\g$, yields the relation 
\begin{equation*}
\frac{D^2\@X^i}{D s^2}=R^{i}_{jkl}\@\d{x^j}/d s\@\d{x^k}/d s\@ X^l= \frac\Lambda 3\,\big(\delta^i_k\@g_{jl}- \delta^i_l\@g_{jk}\big)\@V^j\@V^k\@X^l=\frac\Lambda 3\,X^i
\end{equation*}

The value of $\Lambda$ is therefore related to the acceleration associated with the expansion of the Universe. Observational data currently appear to rule out the choice 
$\Lambda<0$. 

Beyond this, eqs.~\eqref{1.13}, \eqref{1.14} entail the relations
\begin{equation}\label{1.15}
V^p_{\;\;\|p}\@=\@3\,\frac{\dot R}R\,,\qquad \DD0\big(V^p_{\;\;\|p}\big)= 3\;\frac{R\@\ddot R-\dot R^2}{R^2}\,= 3\,\frac k{R^2}
\end{equation}
which, combined with eq.~(\ref{1.12}a), single out $k = -1$ as the only admissible value.
 
Substituting this into eq.~(\ref{1.14}b), and integrating under the initial condition $R(0) = 0$, we conclude that the only Friedmann–Robertson–Walker geometry 
consistent with both the Machian perspective and the available observational data is uniquely determined by 
\begin{equation}\label{1.16}
k=-1\@,\qquad R\/(t)\@=\@T\@\sinh\frac t T                                                                                    
\end{equation}
with $T = \sqrt{\Frac{3}{\Lambda}}$ reflecting the acceleration associated with the expansion of the Universe.\vspace{1pt}

In the limit $T \to \infty$ (non-accelerated expansion), eq.~\eqref{1.16} reduces to
\begin{equation}\label{1.17} 
k=-1\@,\qquad R\/(t)\@=\@t                                                                                    
\end{equation}
corresponding to Milne's cosmological model

\medskip\noindent
(iii) To complete the discussion of the cosmological problem, we still have at our disposal the pair of equations (\ref{1.12}a,b) and the continuity 
equation\eqref{1.10}. 

In view of eq.~\eqref{1.15}, the latter can be rewritten as 
\begin{equation}\label{1.18}
0=\DD0(\mu)+3\@\mu\,\frac{\dot R}R\,=\@\mu\,\DD0\log\big(\mu\@R^3\big) \;\Longrightarrow\;\mu\?R\?^3\coloneq M\@=\const
\end{equation}

When eqs.~\eqref{1.15} and~\eqref{1.18} are inserted into eqs.~(\ref{1.12}a,b), they determine the evolution laws for the coupling coefficient $\Chi$ and the pressure 
$p$, namely, 
\begin{subequations}\label{1.19}
\begin{align}
& \Chi\@=\@\frac3{\mu\/R\?^2}\,=\,\frac{3\@R}M\,,                                                                        \\[3pt]
& p\@=\@\frac23\,\a\@\mu\@V^p_{\;\;\|p}\@=\@2\@\a\@\mu\,\frac{\dot R}R
\end{align}
\end{subequations}

If we interpret $R(t)$ as an estimate of the size of the Universe at cosmic time~$t$, eq.~(\ref{1.19}a) is consistent with the results of Thirring and Lense regarding 
the relation between the radius of the Universe, its material content, and what they considered as the ``gravitational constant'' $\Chi$ \cite{Moller,Sciama}. 

Eqs.~\eqref{1.9}, \eqref{1.19} suggest the following evolutionary scenario:
\begin{itemize}
\item
An early phase characterized by an effective equation of state $p>\mu/3$, in which the cosmological substratum behaves as a pressure-dominated primordial fluid. This 
regime is assumed to hold immediately after the Big Bang (and any putative primordial smoothing/inflationary stage), and it persists as long as 
\[
\@\a\,\Frac{\dot R}{R}\,>\,\Frac16\,.
\]
By virtue of eq.~\eqref{1.16}, this condition breaks down at the instant $t_0$ determined by 
\begin{equation}\label{1.20}
T\@\tanh\frac{t_0}{T} \@=\@ 6\@\a
\end{equation}
($\@t_0=6\@\a \text{ in the limit } T\to\infty,\ \text{i.e. for the Milne model}\@$). 

In a standard $\Lambda$CDM calibration, it is natural to identify $t_0$ with the epoch of photon decoupling (surface of last scattering), whose redshift is 
$z_\ast\simeq 1090$ and whose cosmic time is $t_0\simeq 3.8\times 10^5\ \mathrm{yr}$ \cite{Planck2018VI,SeagerSasselovScott2000}. 

\item
For $t>t_0$, the tensor~\eqref{1.9} can be decomposed as 
\begin{equation}\label{1.21}
\th_{ij}\@=\@\bar\mu\@V_i\@V_j\@+\@\rho\Bigl(V_i\@V_j\@+\Frac13\,\g_{ij}\Bigr),
\qquad
\rho=3\@p,\qquad
\bar\mu=\mu-\rho\,,
\end{equation}
so that the substratum effectively splits into a (pressureless) matter component of density $\bar\mu$, with matter tensor $\bar\mu\,V_iV_j$, and a radiative component 
(photon gas) represented by $\rho\bigl(V_iV_j+\frac13\,\g_{ij}\bigr)$. 

It is only after $t_0$ that the Universe becomes electromagnetically transparent, so that all photon-based observational probes (from the CMB to galaxy surveys) can 
constrain, at best, the interval $(t_0,t)$ between decoupling and the present cosmic time $t$. 
\end{itemize}

A noteworthy consequence concerns the cosmic background radiation, which encodes the state of the radiative component at (and shortly after) $t=t_0$. 

Presently, the radiative energy density is extremely small compared with the dominant components: using the measured CMB temperature 
\linebreak
$T_0=2.72548\pm0.00057\ \mathrm{K}$ \cite{Fixsen2009,Planck2018VI}, 
one finds that the photon fraction is of order $10^{-5}$--$10^{-4}$, so that a representative estimate can be written as 
\[
\frac{\rho}{\hat\mu}\approx 6\times 10^{-5},\qquad\text{hence also}\qquad \frac{\rho}{\mu}=\frac{\rho}{\rho+\hat\mu}\approx \frac{\rho}{\hat\mu}\,.
\]

On the other hand, from eqs.~(\ref{1.19}b) and~\eqref{1.20} we obtain 
\begin{equation}\label{1.22}
\frac{\rho}{\mu}=\frac{3\@p}{\mu}=6\@\a\,\frac{\dot R}{R}=\tanh\!\left(\frac{t_0}{T}\right)\bigg/\tanh\!\left(\frac{t}{T}\right)\!.
\end{equation}

Taking $t_0\simeq 3.8\times 10^5\ \mathrm{yr}$ and the present cosmic time $t\simeq 13.8\times 10^9\ \mathrm{yr}$ (Planck reports $t_0^{\rm today}=13.797\pm0.023\ 
\mathrm{Gyr}$ in the base-$\Lambda$CDM analysis \cite{Planck2018VI}), eq.~\eqref{1.22} yields an implicit estimate of the parameter $T$, and therefore of the 
cosmological constant via $\Lambda=3/T^2$. A direct evaluation gives 
\begin{align*}
& T = 6.52 \times 10^9 \;\text{years}\@,                                                                                                            \\[3pt]
& \Lambda = 7.057 \times 10^{-20} \;\text{years}^{-2} \quad\bigl(= 7.884 \times 10^{-52} \;\text{m}^{-2}\ \text{in MKS units}\bigr)\@.
\end{align*}
For comparison, the $\Lambda$CDM fit to Planck data corresponds to a cosmological constant of order $\Lambda \sim 10^{-52}\ \mathrm{m}^{-2}$ \cite{Planck2018VI}, i.e. on 
the same characteristic scale as the value inferred above.

\subsection{Concluding remarks}\label{S1_4}
The availability of a global relationship between the material distribution and the geometry of the Universe, such as the one expressed by the equation
\begin{equation}\label{1.23}
G_{ij}+\Lambda\@g_{ij}-\DD0\big(V^p_{\;\;\|p}\big)\Big[V_i\@V_j+\a\big(V_{i;j}+V_{j;i}\big)\Big]=\Chi\@\th_{ij}\@,
\end{equation}
is appealing both conceptually and aesthetically.

However, when addressing the case of the actual Universe, the application of Eq.~\eqref{1.23} raises certain nontrivial foundational issues. 

In this context, the unknowns include, in addition to $\DD0$ and $\Phi$, the dynamical fields specifying the matter tensor $\th_{ij}$. 

By supplementing Eq.~\eqref{1.23} with a suitable system of constitutive and/or evolution equations, one might hope to determine all unknowns by formulating a well-posed 
Cauchy problem. 

As noted in Sec.~\ref{S1_2}, however, this expectation clashes with a fundamental conceptual issue: the Universe, being ``the whole'', can only be said to evolve with 
respect to itself. Any statement regarding its evolution would therefore be tautological. 

The concepts of momentum and energy, as well as the corresponding evolution laws, are in fact meaningful only within the framework of a local physical theory, which 
implicitly assumes the presence of an external background to refer to. 

This does not rule out the possibility of describing the cosmic matter distribution by means of a tensor $\th_{ij}$, but it does preclude interpreting $\th_{ij}$ as an 
energy-momentum tensor in the proper sense of the term. 

What does make sense, rather, is to analyze the dynamical behavior of the inhomogeneities present in the real Universe.

This is precisely the line of approach adopted in Sec.~\ref{S1_2}, which renounces any deterministic claim concerning the Universe as a whole, treats the field $\DD0$ as 
a hidden variable, and reformulates eq.~\eqref{1.23} in the form given by\eqref{1.7}. 

Of course, the tensor $T_{ij}$ on the right-hand side of eq.~\eqref{1.7} no longer represents the actual matter distribution. Rather, recalling that, in the absence of 
local inhomogeneities, the cosmological background satisfies eq.~\eqref{1.11}, it can be interpreted as a description of how the real distribution departs from the 
smoothed-out one --- that is, as a representation of matter overdensities and/or underdensities with respect to the average value. 

This, once again, supports interpreting eq.~\eqref{1.7} as an Einstein-like equation, not for the overall bulk of matter, but for its inhomogeneities. 

In this connection, the fact that the coupling coefficient~$\Chi$ is not a constant but rather a scalar field on~$\V4$ warrants further consideration. Treating it as an 
unknown would necessitate the introduction of an additional scalar equation. Assuming that the energy-momentum tensor $T_{ij}$ can be expressed --- via appropriate 
equations of state --- in terms of four variables (typically, the density and four-velocity of the inhomogeneities), eqs.~\eqref{1.7} do in fact provide ten equations 
for eleven unknowns. 

A more convenient approach is to regard $\Chi$ as a function of the state of the Universe as a whole, thus situating its evaluation within a cosmological framework. From 
this perspective, one can take advantage of the fact that, according to eq.~(\ref{1.19}a), the function $\Chi$ is a \emph{first integral\/} of the distribution~$\mathcal 
D$ --- more specifically, that the differential $d\?\Chi$ is an everywhere nonzero 1-form, vanishing along each hypersurface~$t = \const$ 

This allows for a refinement of the notion of the ``footprint left by the reference background'', by identifying it with the field~$\Chi$ and treating it as an intrinsic 
attribute of~$\V4$. In this way, $\Chi$ ceases to be an unknown and instead enters the gravitational equations as a datum. 

The operational scheme emerging from the above discussion can be outlined as follows.

The primary element to focus on is the presence of inhomogeneities in the matter distribution and how, through suitable constitutive equations, their description can 
ultimately be reduced to specifying an energy-momentum tensor $T_{ij}$ depending on four independent thermokinetic variables as well as on the metric. 

At this stage, the properties of the overall Universe are formally disregarded. Not so its presence, whose imprint is captured by the Riemannian hypothesis stated in 
Axiom~\ref{Ax1}. In particular, the matter tensor $\th_{ij}$ is set aside and treated as one of the unknowns of the problem. 

Eqs.~\eqref{1.23} are then recast in the form \eqref{1.7}, with $\Chi$ regarded as known, and employed --- possibly in conjunction with suitable initial and boundary 
data --- to determine the metric tensor and the evolution of the inhomogeneities. 

Once this has been accomplished, the four-velocity of the substratum is expressed by the relation 
\begin{equation}\label{1.24}
\DD0\@=\@\frac{-\@g^{ij}\@\Chi_{,\?j}}{\sqrt{\mathstrut -g^{pq}\@\Chi_{,\?p}\@\Chi_{,\?q}}}\;\de{}/de{x^i}
\end{equation}
while the matter tensor~$\th_{ij}$ is obtained by inserting all results into eq.~\eqref{1.23}.

The evaluation of the fields $\DD0$ and $\th_{ij}$ is thus a purely computational task, which in most cases may even be regarded as dispensable.

The separation between the study of the geometrical environment determined by inhomogeneities and the subsequent reconstruction of the matter distribution in the 
surrounding Universe proves particularly fruitful in local applications of the theory. 

When the variations of the function $\Chi$ in the region under consideration are sufficiently small, one may reasonably neglect them, setting $\Chi = \text{const}$ and 
treating eq.~\eqref{1.7} as an ordinary Einstein equation. 

While perfectly legitimate, this choice suppresses the role of $\Chi$ as a marker of the reference background, thereby ruling out any possibility of reconstructing the 
global cosmological picture through eqs.~\eqref{1.23} and \eqref{1.24}, and reaffirming instead the purely local character of the procedure. 

If properly interpreted, the previous discussion qualifies General Relativity as the local counterpart of a fully Machian theory. 

Indeed, owing to the extremely small present value of the average density $\mu$ of the Universe, placing the origin of gravitation not in the matter distribution itself 
but rather in its inhomogeneities has no appreciable effect on the numerical outcome of the local equations. 

Significant differences, however, are expected to arise in astrophysical applications, where the presence of matter concentrations must necessarily be accompanied by 
corresponding rarefactions, which --- within a Machian framework --- should be interpreted as negative sources of gravitation. 

This point, which may prove relevant to the explanation of phenomena commonly attributed to the presence of dark matter, will not be pursued further in the present 
paper.

\section{Dynamics}\label{S2}

\subsection{Temporal Variation of Fundamental Constants in Physics}\label{S2_1}
The dependence of the coefficient $\Chi$ on the age of the Universe, as outlined in Eq.~(\ref{1.19}a), has far-reaching implications for all other physical 
``constants''. 

To allow for all possible scenarios, it is convenient --- at least at a preliminary stage --- to refrain from adopting natural units and to explicitly include the speed 
of light where appropriate. 

To begin with, we observe that the evolution law for a freely falling point particle, obtained by applying the Einstein-like equation \eqref{1.7} along with the 
Lichnerowicz matching conditions to an incoherent dust filling a spatially infinitesimal world tube, takes the form 
\begin{equation}\label{2.1}
\dD{}/d{\tau}\big(\Chi\@m\/V^i\big)\@=0
\end{equation}

This implies that the resulting world-line is geodesic, and that it is not the particle’s rest mass $m$ that is conserved, but rather the product $\Chi m$. The masses of 
bodies are thus influenced by the expansion of the Universe, acquiring a time dependence scaling as 
\begin{equation}\label{2.2}
m\sim\Chi^{-1}
\end{equation}

From a Machian viewpoint, this is hardly surprising: it simply quantifies the fact that the progressive recession of cosmic matter reduces its contribution to local 
inertia --- and hence the inertial mass of bodies. 

Recalling that, with the normalization $g_{ij}V^i\@V^j=-1\@$ adopted so far, $\Chi$ is related to Newton's gravitational constant $\g$ by the equation 
\begin{equation*}
\g = \frac{\Chi\@c^2}{8\@\pi}\,,
\end{equation*}
we may draw the following conclusions:
\begin{itemize}
\item
The gravitational attraction between two point particles at a fixed distance is described by a force whose magnitude exhibits a time dependence of the form
\begin{equation*}
F\sim \g\@m^2\sim c^2\@\Chi\@m^2\sim \frac{c^2}\Chi
\end{equation*}
If we accept as plausible the assumption that the balance of local forces is not affected by variations in inertia, we are led to conclude that all other local 
interactions must exhibit the same behavior. 

In particular, on the basis of Coulomb's law, and working in Gaussian units, the electric charge $q$ satisfies the relation 
\begin{equation}\label{2.3}
q^2\sim \frac{c^2}\Chi\,.
\end{equation}
Together with eq.~\eqref{2.2}, this shows that the classical radius of the electron, expressed in terms of the charge $e$ and the mass $m_e$ as
\begin{equation*}
r_e=\frac{e^2}{m_e\@c^2}
\end{equation*}
is independent of the age of the Universe. 

\item
The Bohr radius of the hydrogen atom is related to the classical electron radius by the expression 
\begin{equation*}
r_a=\frac{r_e}{\a\?^2}
\end{equation*}
where $\a=\Frac{e^2}{\hbar\@c}$ denotes the fine-structure constant.\vspace{-2pt}

In order to preserve the ratio $\Frac{r_a}{r_e}$\vspace{2pt}, the coefficient $\a$ must remain constant throughout the evolution of the cosmos. It follows that 
Planck's ``constant'' must satisfy the evolution law 
\begin{equation}\label{2.4}
h \sim \frac{e^2}{c} \sim \frac{c}{\Chi}\,.
\end{equation}

\item
The atomic energy levels behave as 
\begin{equation}\label{2.5}
E_n\@=\@-\@\frac{2\@\pi^2\@Z^2}{n^2}\,\frac{m\@e^4}{\hbar^2}\sim m\@c^2\sim \frac{c^2}\Chi\,.
\end{equation}
The emitted frequencies scale as $\nu\propto E_n/h\sim c\@$. Hence, if these frequencies are to remain independent of the emission time --- which is essential for 
using the galactic red-shift as a measure of distance --- $c$~must remain constant. With this choice, atomic clocks measure time intervals that are unaffected by the 
expansion of the Universe. 

The same conclusion applies to a pendulum on Earth's surface, whose frequency $\@\nu=\Frac1{2\@\pi}\@\sqrt{\Frac g\ell}\@$ behaves like $\sqrt g= 
\sqrt{\Frac{\g\@M_T}{R_T^2}}\sim c$.
\end{itemize}

\subsection{Particle and light rays dynamics}\label{S2_2}
The arguments of Sec.~\ref{S2_1} call for a partial revision of the laws of dynamics.

We retain the assignment of a four-momentum $P=m\@V$ to each material particle, requiring that, in any collision, the vector sum of the incoming four-momenta equals that 
of the outgoing ones, while keeping in mind that this statement is purely local and does not entail global conservation of four-momentum. 

According to eq.~\eqref{2.1}, the evolution of a freely falling point particle $\p$ satisfies
\begin{equation}\label{2.6}
\dD{}/d{\tau}\big(\Chi\?P\big)\@=\@\Chi\@\biggl(\dD{P}/d\tau +\d{\log \Chi}/d\tau\,P\biggr)=\@0\@,
\end{equation}
so that it is the product $\Chi\@P$, rather than $P$ itself, that is conserved.

If one maintains the usual identification of the absolute derivative $\Frac{DP}{D\tau}$ with the four-force $F$, eq.~\eqref{2.6} implies that the cosmic background 
generates a universal effect given by $-\D{\log \Chi}/d\tau\,P$.\vspace{3pt} Accordingly, in the presence of other interactions, the equation of motion reads 
\begin{equation}\label{2.7}
\dD{\?P}/d{\tau}\@=\@F\@-\@\d{\log \Chi}/d\tau\,P
\end{equation}

For ordinary material particles, the ``cosmological'' term is essentially unobservable, as its detection would require monitoring over timescales long enough for 
significant variations in $\Chi$ to occur. 

A scenario more accessible to experimental verification concerns the behavior of photons.
In this case, the four-momentum is related to the wave four-vector $k=k^i\@\De{}/de{x^i}$\vspace{3pt} via the relation $P=h\@k$. In view of eq.~\eqref{2.4}, the 
conservation of $\Chi\?P$ then implies the conservation of $k$. 

This property has an immediate application in the analysis of galactic red-shift. In the rest frame of the material substratum, let $\g$ denote the world-line of a light 
ray and $k$ its associated wave four-vector. 

We parameterize $\g$ as $x^i = x^i(\xi)$, so that the wave four-vector can be identified with the tangent vector $k =\D\g/d\xi\@$, whence in particular 
\begin{equation*}
\d t/d\xi\,=k^0=\@\nu\@.
\end{equation*}

In this setting, on the one hand, we have the equation
\begin{equation*}
\dD{\?k}/d\xi\,=\@k^i_{\;\@;\@j}\,k^j\@=\@0 
\end{equation*}
which expresses the geodesic character of $\g$ and the affine nature of the adopted parametrization. 

On the other hand, by the very definition of $k$, the frequency of the ray in the rest frame of the substratum is given by the scalar product 
\begin{equation*}
\nu\@=-\big(k\@,\DD0\big)\@=\@-\,k^i\@V_i
\end{equation*}

Reparameterizing $\g$ in terms of $t = t\/(\xi)$, and using eqs.~(\ref{1.12}c), \eqref{1.15}, along with the identification $\D{\?t}/d\xi=\nu$, yields the evolution law: 
\begin{align*}
\d\nu/d t&=-\@\frac1\nu\biggl[\biggl(\cancel{\dD k/d\xi}\@,\@\DD0\biggr)+ \biggl(k\@,\@\dD{\DD0}/d\xi\biggr)\biggr]=                      \\[3pt]
&=-\@\frac1\nu\,k^i\@V_{i\?;\?j}\,k^j= -\@\frac{V^p_{\;\;\,;\@p}}{3\@\nu}\;\g_{ij}\,k^i\@k^j=-\@\frac{\dot R}R\,\@\nu
\end{align*}

Denoting by $\nu_e\@$ the frequency at emission time $t_e\@$ and by $\nu_o\@$ the frequency at observation time $t_o\@$, the red-shift formula is thus recovered:
\begin{equation*}
\d{}/dt\,\log(\nu\@R)=0\quad\Longrightarrow\quad \nu\@R=\const \quad\Longrightarrow\quad \nu_e\@R\/(t_e)=\nu_o\@R\/(t_o)\@.
\end{equation*}

\section*{Conclusions}
In this article we have recast Mach’s principle in a genuinely geometric form and used it as the organizing idea of our discussion. 

From the Machian standpoint, the motion of any system must be interpreted as the net outcome of its interaction with the rest of the Universe: the total dynamical effect 
acting on a test particle depends on the particle’s own attributes, on the state of the external Universe, and on the relations between them, with the evolution 
determined by an overall balance condition. 

Within a pseudo-Riemannian setting (Axiom~\ref{Ax1}), this viewpoint raises a sharp conceptual question: is geodesic motion merely \emph{selected} by local field 
equations on a pre-assigned arena, or is the very \emph{existence} and global form of the arena --- namely the full pseudo-Riemannian structure of $\V4$ --- to be 
attributed to the Universe as a whole? 

The Machian notion of free fall --- and, more generally, inertial motion --- presupposes the existence of a background relative to which such motion is defined. 

For genuinely \emph{local} systems this requirement is met by the presence of the rest of the Universe: distant matter provides the analogue of Newton’s ``fixed stars'', 
while its influence is encoded in the space-time geometry used to define inertial frames and local dynamics. 

In this regime, General Relativity remains empirically unsurpassed and conceptually compelling. Moreover, as stressed in Proposition~\ref{Pro1_1}, the absolute and 
Machian readings of local applications lead to the same operational conclusions; any choice between them must therefore be justified on cosmological grounds, rather than 
by solar-system phenomenology. 

The situation changes radically when the ``system'' under investigation is the Universe itself. 
In cosmology there is no external environment left to provide the background needed to define inertial motion. The origin of the relevant space-time structures must 
therefore be traced back to the Universe itself, i.e., to the object under study. 

This motivates the Machian standpoint adopted here: the material content of the Universe is not primarily responsible for \emph{differential} gravitational effects on 
top of a pre-existing arena, but for the \emph{existence and global form} of the pseudo-Riemannian structure of $\V4$ itself. 

Moreover, the explicit metric structure is taken to depend essentially on the \emph{distribution} of matter, with gravitation associated with inhomogeneity, and with the 
homogeneous/isotropic limit corresponding to a maximally symmetric (constant-curvature) geometry. 

Within this perspective, Einstein’s equations can be retained in the local regime while being reinterpreted cosmologically: the tensor $T_{ij}$ entering 
\[
G_{ij}+\Lambda g_{ij}=\Chi\,T_{ij}
\]
is naturally interpreted as encoding the mechanical content of the \emph{inhomogeneities} of the actual material distribution, rather than the total matter content of 
the Universe. 

It is useful to emphasize how this program differs from other prominent Mach-inspired approaches. 

In Sciama’s inertial-induction picture, the focus is on a dynamical mechanism whereby distant matter induces local inertia, typically formulated in a 
quasi-field-theoretic manner \cite{Sciama}. 

In scalar--tensor gravity, historically initiated by Brans and Dicke, Machian content is implemented by enlarging the gravitational sector, so that additional long-range 
degrees of freedom mediate ``cosmic influence'' (often through an effective, dynamical gravitational coupling) \cite{BransDicke1961}. 

While scalar--tensor frameworks have since been vastly generalized and systematized --- from covariant classifications of healthy single-field models 
\cite{Deffayet2011,Kobayashi2019HorndeskiReview,Gleyzes2015BeyondHorndeski,LangloisNoui2016Degenerate,LangloisNouiRoussille2021QuadraticDHOST} to effective descriptions 
of cosmological perturbations \cite{Gubitosi2013EFTDE,BelliniSawicki2014Alpha,FrusciantePerenon2020EFTReview} --- their Machian thrust typically proceeds by introducing 
(and then constraining) extra gravitational degrees of freedom. 

Similarly, the Hoyle--Narlikar theory incorporates Machian ideas by embedding global matter influence directly into the postulates of the gravitational theory 
\cite{HoyleNarlikar1964}, and relational programs emphasize the elimination of absolute structures at the level of the dynamical laws 
\cite{BarbourBertotti1982,Barbour2010}. 

By contrast, the Machian program pursued here shifts the logical emphasis: in cosmology, the key issue is not merely how dynamics is modified \emph{on} a background, but 
how the background structures that render inertial concepts meaningful are to be attributed to the Universe itself. 

This perspective makes the global specification of the cosmological model essential, since inertial structure cannot be fixed locally alone. 

Operationally, it requires defining a smoothing or averaging procedure that separates the large-scale background geometry from the inhomogeneous matter distribution. 

If gravitation is tied to inhomogeneity on top of a maximally symmetric limit, one must clarify (i) what is meant by ``smoothing out'' the Universe and (ii) how the 
effective geometric structures arising from such a smoothing relate to the exact geometry sourced by the actual distribution. 

These analyses should illuminate how --- and in what sense --- global matter distributions can determine local inertial frames within General Relativity, in the choice 
of global conditions rather than through local field equations alone \cite{LyndenBellKatzBicak1995,BondiSamuel1997,Schmid2006,Schmid2009}. 

Taken together, these considerations sharpen the core message of this article. Reconsidering the role of Einstein’s equations in cosmology and reformulating the 
cosmological problem so that the distribution of matter is responsible first and foremost for the very space-time structures that make inertial concepts meaningful also 
reconnects with the historical trajectory of Einstein’s own Machian motivations and the subsequent debates on their precise implementation 
\cite{Einstein1918,Hoefer1994,BarbourPfister1995,LichteneggerMashhoon2004}.

\bibliography{refs}

\end{document}